\documentstyle[epsf,psfig,epic,eepic,fortschritte]{article}
\def\titleline{On the Heisenberg principle, namely on the
information-disturbance trade-off in a quantum measurement}
\def\authors{Giacomo Mauro D'Ariano}
\def\addresses{{\em Quantum Optics and Information Group},
Istituto Nazionale di Fisica della Materia, Unit\`a di Pavia\\
Dipartimento di Fisica ``A. Volta'', via A. Bassi 6, I-27100 Pavia, Italy\\
{\tt www.qubit.it}}
\def\abstracttext{Common misconceptions on the Heisenberg principle
are reviewed, and the original spirit of the principle is reestablished in
terms of the trade-off between information retrieved by a measurement
and disturbance on the measured system. After analyzing the
possibility of probabilistically reversible measurements, along with
erasure of information and undoing of disturbance, general
information-disturbance trade-offs are presented, where the disturbance
of the measurement is related to the possibility in principle of
undoing its effect.}
\def\operatorname#1{\mbox{#1}}
\def\<{\langle}\def\>{\rangle}
\def\Tr{\operatorname{Tr}}\def\:{\hbox{\bf :}}
\def\){\rangle\!\rangle}\def\({\langle\!\langle}

\def\set#1{{\sf #1}}
\def\transp#1{{#1}^\tau}

\def\rnk{\operatorname{r}}
\def\Rng{\set{Rng}}\def\Ker{\set{Ker}}

\def\diag{\operatorname{diag}}\def\dim{\operatorname{dim}}

\def\l2{{\mathbf l}_2}

\def\N#1{|\!|#1|\!|}

\def\sH{\set{H}}
 
\def\Ens{{\cal E}}\def\ap{{\mathbf a}}\def\z{{\mathbf z}}
\def\eg{e. g. }\def\ie{i. e. }
\begin{document}
\makefront
\section{Introduction} 
The need for hard miniaturization and the recent discovery of
radically new information processing\cite{nielsenbook}, have
dramatically changed our attitude towards Quantum Mechanics, 
which eventually got out the middle age of purely academical
consideration, to become a relevant chapter of the modern information 
technology.  At the beginning ``quantum'' was a synonymous of
``uncertainty'', and was considered just as a major limitation in 
nanotechnology. More recently, however, we learned how to turn the
``quantum'' into a powerful horse that we can harness and ride, with
unimagined possibilities in principle for guaranteed cryptographic
communications and tremendous speedup of complex computational 
tasks, giving birth to the new quantum information technology.
\par In the theoretical research for quantum information, one of the
main programs is undoubtedly to establish the actual 
limitations and controllability of quantum measurements, in a unified 
framework suited to the needs for optimization and engineering.
However, looked with not expert eyes, this program should appear quite
incompatible with the paradigm itself of quantum mechanics: the
so-called ``Heisenberg principle'', which establishes the
``participatory'' nature of the quantum experiment. In fact, according
to its popular version---based on the {\em gedanken}
experiment of the $\gamma$-ray microscope
\cite{Heisenberg,Heisenberg_book}, which was 
then elevated to ``principle'' by Ruark \cite{Ruark}---it is impossible
to measure one variable, say the momentum $p$, of a conjugated pair
(\eg position $q$ and momentum $p$) without ``disturbing'' the value of
the conjugated variable $q$ of an amount $\Delta q$ no less than the
order of $\hbar/\Delta p$, where $\Delta p$ is the accuracy of the
measurement \cite{vonneuman}. And such paradigm is not just  
a folklore for the layman, since the principle is clearly stated and
emphasized in excellent textbooks of quantum mechanics---\eg the
valuable Messiah book \cite{Messiah}, which devotes a lengthy section
to the  ``uncontrollable disturbance during the operation of  measurement'', 
with an extensive analysis of different thought experiments in support
of the generality of the ``principle'', and concluding that ``the
unpredictable and uncontrollable disturbance suffered by
the physical system during a measurement is always sufficiently strong
that the uncertainty relations always hold true.''  Misunderstanding
and misuses even at the level of advanced research are revealed, for
example, by the controversy \cite{Braginski,Caves,Yuen_contractive,
Lynch,Caves-defense,Ozawa_SQL_break,Ozawa_SQL_break2,
Ozawa_model_SQL,Yuen_SQL_unpub} 
on the existence of a ``standard quantum limit'' for precision in
monitoring a free mass position---a problem which arose in the
field of gravitational wave detection. Finally, the controversial
nature of the Heisenberg principle is also witnessed by the existence
of an entire book on quantum measurements\cite{Braginski_book} based
on the use of the principle beyond its original heuristic nature, in
contrast to some ``classics'' of quantum mechanics that not even
mention it---\eg the Landau and Lifshitz 
book \cite{Landau}---whereas, for example, if you look for
``uncertainty principle'' in the subject index of the Peres
book \cite{Peres} the referred page number is provocatively the page of
the index entry itself.  
\par Before proceeding with the discussion on the Heisenberg
principle, let me first clarify some common confusion between
``uncertainty relations'' and ``uncertainty principle'', the former
concerning the statistics of repeated measurements on an ensemble of
equally prepared identical quantum systems, the latter, on the
contrary, concerning a sequence of measurements on the same 
quantum system (this difference is well emphasized in the Jammer book
\cite{Jammer2}). The ``uncertainty relations'' do not have any bearing
on the issue of the measurement disturbance, since it can be
experimentally tested by measuring each of the observables
separately: at most one of the two root mean squares, say
$\Delta p$ can be considered as the precision of the {\em
preparation}, \eg by a collimator of particle momentum, and
then $\Delta q$  will results from the statistics of measuring only
$q$. In other words, both $\Delta p$ and $\Delta q$ are {\em a
priori} uncertainties according to the Born rule, and neither will
result as a consequence of the disturbance due to the measurement. As
a matter of fact, since both $\Delta p$ and $\Delta q$ are intrinsical to
the wave function before the measurement, they cannot be logically
connected to the interaction with apparatus. And in fact, a
measurement model was provocatively proposed by Ozawa
\cite{Ozawa_reduct} in which the position of the particle can be measured
leaving it in a eigenstate of the momentum. With no proper
distinction between preparation and measurement (this issue is
extensively analyzed in the recent paper by Muynck \cite{Muynck}) the
two forms of complementarity amalgamated, leading to another erroneous
interpretation of the Heisenberg principle as related to
{\em joint} measurements (see for example the Bohm book \cite{Bohm}). 
Although in quantum mechanics of Dirac and von Neumann joint
measurement of only compatible observables are allowed---in full
logical contradiction with the last interpretation---however, there
are precise indirect models \cite{ArthursKelly,GordonLouisell}
describing approximate joint measurements (which are actually achieved
in a heterodyne apparatus \cite{Yuen_joint}), and the resulting minimum
uncertainty product in principle is double than the Heisenberg bound
\cite{Yuen_joint, Goodman}---the socalled 3dB of added noise in the
optimal joint measurement. 
\par There is an extensive literature on the various
misinterpretations of the principle, starting since from the origins.
Bohr himself disagreed with Heisenberg on the gedanken experiment of
the $\gamma$-ray microscope, as quoted in the original paper
\cite{Heisenberg}. Lamb \cite{Lamb} criticized the 
$\gamma$-ray microscope as unsuitable for position measurements. 
Historical reviews can be found, for example, in the Jammer books
\cite{Jammer1, Jammer2}, and in the Beller book \cite{Beller}.
A serious criticism to the use of the classical definition of
resolving power due to diffraction in the gedanken 
experiment is made in Ref. \cite{Roychoudhury}, where Heisenberg's 
microscopes with super-resolutions violating the principle are
devised. Criticisms to the use of root mean square as measures of
uncertainty and disturbance are made in various papers (see, for
example, Ref. \cite{Uffink}). As regards the ``uncertainty
relations'', there have been many alternative derivations and
generalization since from the origins (see Ref. \cite{Jammer1} for a
detailed history). The general formulation for any pair of non
commuting observables is due to Robertson \cite{Robertson}, after some
relevant remarks of Condon \cite{Condon}. Schr\H{o}edinger
\cite{Schroedinger} then recognized that the uncertainty product is
not invariant under unitary transformations, and found ``tighter''
uncertainty relations. More recently ``entropic'' generalizations of
the uncertainty relations were given \cite{Deutsch,Massen},
noise-dependent relations in Ref. \cite{Hall}, higher-order
uncertainty relations also involving more than two operators
\cite{Sudarshan}, only to quote some work known to the present author.
\par Coming back to the original problem of the Heisenberg gedanken
experiment, even though it is clear that the ``uncertainty relations''
do not have any bearing on the issue of the measurement disturbance,
and there is no in-principle ``uncontrollable disturbance during the
operation of  measurement'',  however, the issue of the minimum
disturbance in-principle from a  quantum measurement in relation with
the information gained from the measurement is still an unsolved
problem.  That a kind of Heisenberg principle must exist in form of
information-disturbance trade-off is evident, for example, from the
impossibility of determining the wave-function of a single system from
any sequence of measurements on the same quantum system
\cite{DArianoYuen_clon}. Such possibility has recently intrigued
several authors \cite{Yamamoto,Aharonov,Kitagawa,Imamoglu,Royer},
which explored concrete measurement schemes based on vanishingly weak
quantum nondemolition measurements \cite{Yamamoto}, weak 
measurements on ``protected'' states \cite{Aharonov}, ``logically
reversible'' \cite{Kitagawa}, and ``physically reversible'' 
\cite{Imamoglu,Royer} measurements. In each of these schemes the 
conclusion is that it is practically impossible to measure the wave
function of a single system, either because the weakness of the
measuring interaction prevents one from gaining information 
on the wave function \cite{Yamamoto}, or because the method of 
protecting the state \cite{Aharonov} actually requires some {\em a 
priori} knowledge on the state (this is suggested in Refs. 
\cite{Royer} and \cite{Yamamoto}), or because quantum measurements can
be physically reverted only with vanishingly small probability of
success \cite{Royer}. The impossibility of determining the
wave-function of a single quantum system is dictated by the no-cloning
theorem \cite{Wootters_clon}, which is just a direct consequence of
unitarity of quantum mechanics \cite{Yuen_clon}. Therefore, as a
consequence of the general laws of quantum mechanics, there must be a
detailed balance between information and disturbance, which makes
impossible to determine the state of a single quantum system from any
sequence of measurements on it. 
\par Despite the relevance of the problem of the information-disturbance
trade-off at the foundational level---although a consequence quantum
laws---very little literature can be found on this issue, maybe due
to the difficulty of the problem. The issue also recently became
of practical relevance for posing general limits in information 
eavesdropping in quantum cryptographic communications. For such purpose,
for example, in Ref. \cite{Fuchs-Peres} Fuchs and Peres analyzed some
trade-offs for the two-state discrimination. A part from this work, 
only few studies are known to the present author: the very
interesting analysis by Fuchs \cite{Fuchs} and by Barnum 
\cite{Barnum}, and, only very recently, a definite result by Banaszek
\cite{Banaszek} on a general trade-off between the quality of a single
state estimation and the fidelity between the input and the output
states of the measurement. Also Ozawa \cite{ozawaunpub} has recently 
proposed a general trade-off, which will be mentioned in more details
in the following. Finally, Belavkin \cite{Belavkin} has given a
Heisenberg principle for continuous measurement of the position in the
framework of filtering theory.
\par In this paper some results will be presented in the attempt to
give general a information-disturbance trade-off which holds for any
quantum measurement. The tradeoff must be valid ``in-principle'',
whence at the single-outcome level, not only in average over outcomes,
as those considered in Refs. 
\cite{Fuchs-Peres,Fuchs,Barnum,Banaszek,ozawaunpub}. Also, since it 
should be valid for a general context, the tradeoff has to be independent
on the particular analytical form of information and disturbance,
which is suited to the particular problem at hand (in the analysis
\cite{Fuchs-Peres,Fuchs,Barnum} the fidelity between input and output
has been considered as a measure of the ``disturbance'').
This requirement of generality has led us to consider trade-offs in
form of majorization ``orderings''\cite{Nielsen,Marshall} between the
conditional probability from the measurement and quantities related to
the measurement effect on the input state, the former being the
variables from which one can evaluate any kind of ``information'', the
latter being the source of the ``disturbance''. The disturbance of the
measurement will be related to the possibility in-principle of undoing
its effect, and for this reason we will previously analyze in general
the occurrence of probabilistically reversible measurements. We will
see that when the measurement effect is undone, also the information
retrieved from it is erased, and from this we will argue that in a
cascade of measurements the disturbance can also be decreased,
however, at the expense of losing the previously gained information.  
The case of measuring an ``observable'' will be analyzed in some
detail. The majorization trade-off will then be applied to the common
case of the mutual information retrieved from the measurement: this
will lead us to a trade-off in a form of a bound tighter than the
Holevo bound \cite{nielsenbook}, with the disturbance in the form of a
Shannon entropy versus the singular values of the measurement
``contraction'' (the operator describing the effect of the single
outcome of the measurement). As we will see, the generality of the
majorization relation turns out to be a weakness when a specific case
of information/disturbance is considered, since it proves the tradeoff
validity in a more limited situation than the actual one, depending on
the relation between the measurement and the ensemble of input states.
Finally, we will see that the disturbance obtained in this way agrees
with the ``decrease of entanglement'' due to the measurement when it
acts locally on an entangled state. 
\section{Information-disturbance trade-offs}
Since we are looking for an in-principle trade-off which should account
for the impossibility of determining the state of a single quantum
system for no {\em a priori} knowledge, we need to consider the
general measurement scenario, in which a sequence of measurements on a
single quantum system is performed, with the possibility of changing
the measuring apparatus at each measuring step, e. g. depending on the
outcome from the previous step. Therefore, our information-disturbance
trade-off must be valid at the single-outcome level, not just in
average over outcomes. Moreover, to be true ``in-principle'', we must
consider a situation of perfect control on the measurement, namely the
apparatus is perfectly known, and we are able to perform any
measurement and any unitary transformation at will, according to the 
rules of quantum mechanics. In the following we will refer to such
in-principle situation as {\em perfect technology}.
\subsection*{Notation} 
Throughout this paper, we will use boldfaced letter and square
brackets to denote arrays/vectors, \eg  ${\mathbf x}=
[x_i]=(x_1,x_2,\ldots)$. For any operator $A$ on 
the Hilbert space $\sH$ with $d=\dim(\sH)$, by $\Ker(A)$ we will denote
the kernel of $A$, by $\Rng(A)$ its range, by $\rnk(A)$ its rank, and
by $P_A$ the orthogonal projector on $\Rng(A)$. We will write the
singular value decomposition of $A$ as $A=X_A \Sigma_A Y_A^\dag$, where
$\Sigma_A=\diag\{\sigma_1(A),\sigma_2(A),\ldots,\sigma_r(A),0,\ldots,0\}$
is the diagonal matrix of singular values of $A$ ordered decreasingly
(including also the vanishing ones), and $X_A$ and $Y_A$ are unitary
operators of left and right eigenvectors respectively. By $\N{A}_p\doteq
[\sum_i\sigma_i(A)^p]^{\frac{1}{p}}$ we will denote the $p$-Shatten
norm of $A$, with $\N{A}_1$ the trace-norm, $\N{A}_2$ the
Hilbert-Schmidt norm, and with $\N{A}\equiv\N{A}_{\infty}$ the usual
operator norm. The symbol $A^\ddag$  will denote the
Moore-Penrose pseudoinverse of $A$, \ie 
$A^\ddag=Y_A\Sigma_A^\ddag X_A^\dag$, with $\Sigma_A^\ddag=
\diag\{\sigma_1^{-1}(A),\sigma_2^{-1}(A),\ldots,\sigma_r^{-1}(A),0,\ldots,0\}$,
\ie $A^\ddag$ is the same as $A^\dag$ but with the inverse of
the non-vanishing singular values. The Moore-Penrose pseudoinverse is
completely characterized by the properties $AA^\ddag A=A$, $A^\ddag
AA^\ddag =A^\ddag$,  $(A^\ddag 
A)^\dag=A^\ddag A$, and $(AA^\ddag)^\dag= AA^\ddag$. It follows that
$P_A=AA^\ddag$ and $P_{A^\dag}=A^\ddag A$. We will denote by
$\Ens=(\set{S},\ap)$ the ensemble of states $\set{S}=\{\psi\}$    
distributed with {\em a priori} probability $\ap=[a(\psi)]$
using the abbreviate notations $\psi\in\Ens$ for
$\psi\in\set{S}(\Ens)$, $\set{S}(\Ens)$ and $\ap(\Ens)$ to denote the
set of states and the probability distribution of the ensemble $\Ens$,
respectively, and $|\Ens|$ the cardinality of 
$\set{S}(\Ens)$. The singleton set with the state $\varphi$  
will be denoted by the state itself $\varphi$. We will call {\em
universal ensemble} the uniform ensemble of all possible (pure) input
states. With $\rho_\Ens=\sum_{\psi\in\Ens}a(\psi)|\psi\>\<\psi|$
we will denote  the {\em a priori} density operator of the ensemble
$\Ens$.  
The Shannon entropy of the probability vector $\ap=[a_i]$ will be
denoted by $H(\ap)\doteq 
-\sum_i a_i\log a_i$ and for the ensemble $\Ens$ we will also write
equivalently $H(\Ens)\equiv H(\ap(\Ens))=-\sum_{\psi\in\Ens} a(\psi)\log a(\psi)$.
Finally we will write
$\Ens=p\Ens_1+(1-p)\Ens_2$ for the union ensemble with
$\set{S}(\Ens)=\set{S}(\Ens_1)\cup\set{S}(\Ens_2)$ in which a state is
picked from $\set{S}(\Ens_1)$ or $\set{S}(\Ens_2)$ with probability
$p$ and $(1-p)$, respectively, corresponding to the density operator
$\rho_\Ens=p\rho_{\Ens_1}+(1-p)\rho_{\Ens_2}$, and write
$\Ens=p\Ens_1\oplus(1-p)\Ens_2$ when
$\set{S}(\Ens_1)\perp\set{S}(\Ens_2)$.  
\subsection{Pure measurements}
A measurement with perfect technology means that we have a precise
quantum description of the apparatus. Such a measurement is {\em
pure}, namely it preserves purity of states. A pure measurement
for a single outcome is described by a {\em contraction}
$M$, namely an operator with bounded norm $\N{M}\le 1$, to guarantee
occurrence probability not greater than unit for any input state. The
output state $|\psi_M\>$ after the measurement and the probability
$p(M|\psi)$ that $M$ occurs on the input state $|\psi\>$ are given by
\begin{equation}
|\psi_M\> =\frac{M|\psi\>}{\N{M\psi}}\;\;\mbox{(state reduction)}, \qquad
p(M|\psi)=\N{M\psi}^2\;\;\mbox{(Born rule)}.
\label{outpsi}
\end{equation}
We will also regard the case of unitary $M$ as a limiting case of
``measurement'', which gives no information on $|\psi\>$, since
$p(M|\psi)=1$ independently on $|\psi\>$. This will also
corresponds to {\em no in-principle disturbance} for any state, since
with perfect technology we can deterministically reverse the effect of
$M$ without knowing $|\psi\>$.  
\subsection{Information from a single measurement outcome}
We can always regard the quantum measurement as a problem of 
discriminating between a set of hypotheses corresponding to an
ensemble $\Ens=(\set{S},\ap)$ of states $\set{S}=\{\psi\}$ 
distributed with {\em a priori} probability $\ap=[a(\psi)]$. 
The Shannon entropy $H(\Ens)$ quantifies our {\em a priori} ``ignorance'' on
which-state of the ensemble. When the outcome corresponding to the
contraction $M$ occurred, then our ignorance is reduced, since now the
{\em a priori} probability distribution $\ap=[a(\psi)]$ is upgraded to the {\em a
posteriori} probability $\ap_M=[a(\psi|M)]$ that the state was
$\psi$ given that we know that $M$ has occurred [the corresponding
ensemble will be denoted by $\Ens_M=(\set{S},\ap_M)$]. The
probability $a(\psi|M)$ is
given by the Bayes rule $a(\psi|M)=a(\psi)P(M|\psi)/p_\Ens(M)$,
where $p_\Ens(M)\doteq\Tr[\rho M^\dag M]$ denotes the overall
occurrence probability for $M$.  The information $\Delta I_{\Ens}(M)$
on which-state $\psi\in\Ens$ gained from the occurrence of $M$ is just the
difference between our ignorances before and after the occurrence of 
$M$, namely  
\begin{equation}
\Delta I_{\Ens}(M)=H(\Ens)-H(\Ens_M)=-\sum_{\psi\in\Ens}a(\psi)\log
a(\psi)+\sum_{\psi\in\Ens}a(\psi|M)\log a(\psi|M).\label{infosingle} 
\end{equation}
\subsection{Knowingly reversible measurements}
We say that the effect of a measurement outcome corresponding to the
contraction $M$ is {\em knowingly reversible on a set 
$\set{S}=\{\psi\}$ of input states} if for any {\em a priori} unknown
input state $\psi\in\set{S}$ we can perform another measurement on the
output state $\psi_M$ of $M$ such that for some outcome---say corresponding
to the contraction $\tilde{M}$---we know for sure that the new output
state is the original $\psi$, for all $\psi\in\set{S}$. In other words, the
contraction $M$ is knowingly reversible on $\set{S}$ if there is
another contraction $\tilde{M}$ such that  
\begin{equation}
\tilde{M}M|\psi\>\propto|\psi\>,\qquad\forall\psi\in\set{S}.\label{rev1}
\end{equation}
This means that with some probability we can {\em undo the effect of
$M$} with another measurement contraction $\tilde{M}$.
The squared modulus of the proportionality constant in
Eq. (\ref{rev1}) is the overall probability of achieving $M$ and
knowingly reversing it with $\tilde{M}$. If $\rnk(M)=d$ ($M$ full
rank), then  $M$ is knowingly reversible for any input 
state, since it is invertible as an operator. It is easy to check
that, apart from an overall phase factor, the most efficient {\em
reversion} $\tilde{M}$ (\ie maximizing the reversing probability on
any input state) is given by 
$\tilde{M}=M^{-1}/\N{M^{-1}}$. In fact, by taking $\tilde{M}=\omega
M^{-1}$, the overall probability of 
achieving $\tilde{M}$ on $|\psi_M\>$ multiplied by the probability
$P(M|\psi)$ of achieving $M$ on $|\psi\>$ is just 
$|\omega|^2$ and the maximum $|\omega|$ in order to have $\tilde{M}$ as
a contraction is $|\omega|=\N{M^{-1}}^{-1}$. For the most efficient
reversion $\tilde{M}$ the probability $p_{rev}$ of reversion is
bounded as $\kappa^{-2}(M)\le p_{rev}\le 1$, with
$\kappa(M)=\N{M}\N{M^{-1}}$ the {\em condition number} of $M$, and the
bounds are achieved by the left vectors of the singular value
decomposition of $M$ corresponding to $\sigma_1(M)$ and
$\sigma_d(M)$, respectively. We see that the smaller the condition
number $\kappa(M)$ of $M$, the higher the chance of reversing
$M$, \ie the ``more reversible'' is $M$. Since the condition number
of an operator gives also an error estimate under small perturbations
of the linear action of the operator\cite{Marshall}, this means that
more reversible is $M$, the more ``amplified'' an input perturbation
will result at the output. Also, notice that the probability
$p(\tilde{M}M|\psi)$ of the cascade of $M$ and its successful
reversion is $p(\tilde{M}M|\psi)=|\omega|^2$, independently on the
input state $|\psi\>$, and for the most efficient reversion is 
$p(\tilde{M}M)=\sigma_d^2(M)\le [\prod_n\sigma_n^2(M)]^{1/d}
\le \frac{1}{d}\N{M}_2^2$. The bound $[\prod_n\sigma_n^2(M)]^{1/d}$
generalizes the Bhattacharyya overlap given in Ref. \cite{Ban} for the
case in which the measurement corresponds to an observable $X$ 
(see subsection \ref{s:obs}).
\par When $M$ is not full rank, \ie $\rnk(M)<d$, it 
is still possible to have situations in which $M$ is knowingly
reversible. The first case is when {\em the set $\set{S}$ is
orthogonally split by $M$}, namely it can be written as the union of
two orthogonal subsets $\set{S}=\set{S}_M^\parallel\oplus\set{S}_M^\perp$ of 
which $\set{S}_M^\perp\subseteq\Ker(M)$ and $\set{S}_M^\parallel\subseteq
\Ker(M)^\perp\equiv\Rng(M^\dag)$. In fact, in this case we know a priori
that $M$ cannot occur on an input state $|\psi\>\in\Ker(M)$, whereas
if $M$ occurred, then $|\psi\>\in\Rng(M^\dag)$, and we can reverse $M$
with some probability using a contraction $\tilde{M}$ such that
$\tilde{M}M\propto P_{M^\dag}$, namely
\begin{equation}
\tilde{M}=\omega M^\ddag +Z(I-P_M),\label{MPK}
\end{equation}
where $Z$ is any complex operator. Since $\tilde{M}$ 
must be itself a contraction, from
$\N{\tilde{M}}=\max\{\omega\N{M^\ddag},\N{Z(I-P_M)}\}$ we obtain  
the general parametrization of the most efficient $\tilde{M}$ (a part from a phase
factor)
\begin{equation}
\tilde{M}=\frac{M^\ddag}{\N{M^\ddag}}+Z(I-P_M),\label{MPK1}
\end{equation}
with $Z(I-P_M)$ a contraction.
\par As regards the case in which the set $\set{S}$ is not
orthogonally split by $M$, the contraction can be knowingly
reversible only in the degenerate situation in which $\set{S}$ is the
disjoint union $\set{S}=\set{S}_M^\perp\cup\varphi$ of  
$\set{S}_M^\perp\subseteq\Ker(M)$ with the single state
$\varphi\not\in\Ker(M)$. Since this case is not very interesting (since
it is essentially equivalent to reverse $M$ only on 
a single state), we will not consider it in the following.
\subsection{Negative informations: undoing a measurement erases its information}
In Ref. \cite{Royer} Royer found an example of knowingly reversible
measurement on a two-dimensional space, and supposed that a sequence
of successfully reverted measurements could be used to determine the
state of single quantum system with some probability, without any {\em
a priori} knowledge of the state. However, thereafter in Ref. \cite{Royer_errata} 
he admitted that in fact this was not true. From Eq. 
(\ref{MPK}) we can easily see that in the most general case in which
we are able to revert a contraction $M$, the probability of achieving
$M$ and then reverting it is given by $|\omega|^2$, independently
on the input state, whence any succession of successfully reverted
measurements provides only the information that the input state
was in $\Rng(M^\dag)$, \eg for an ensemble $\Ens$ orthogonally split
by $M$ as $\Ens=p\Ens_M^\parallel\oplus(1-p)\Ens_M^\perp$ 
such information would be
\begin{equation}
\Delta I_{\Ens} (\tilde{M} M)=H(\Ens)-H(\Ens_M^\parallel).\label{revinf}
\end{equation}
For uniform $\Ens_M^\parallel$ Eq. (\ref{revinf}) gives $\Delta I_{\Ens}
(\tilde{M}M)=H(\Ens)-\log(|\Ens^\parallel|)$, and for uniform $\Ens$
one has $\Delta I_{\Ens} (\tilde{M} M)=-\log
p=\log(|\Ens|/|\Ens_M^\parallel|)$. 
For the input universal ensemble necessarily $M$ is
reversible only if $\Rng(M^\dag)\equiv\sH$, and the 
information (\ref{revinf}) is then exactly zero. Since the occurrence
of $M$ must have given some information on which-state of $\Ens$
anyway, this means that undoing the measurement must also erase the
information from it. In fact, the information from a single
measurement outcome in Eq. (\ref{infosingle}) can be negative: the
reader unfamiliar with negative informations should notice that 
the informations considered in the literature are always positive,
since they are averaged over all outcomes, whereas generally the
contribution from a single outcome can be negative.  What 
does it mean to have a negative information? From
Eqs. (\ref{infosingle}) we see that  
negative informations occur when the {\em a posteriori} probability
distribution $\ap_M=[a(\psi|M)]$ is less ``peaked'' around some $\psi\in\set{S}$
than the {\em a priori} probability $\ap=[a(\psi)]$.  In practice, this
means that the measurement result {\em contradicts} our previous
knowledge (see the amusing example by Uffink quoted in the Peres book
\cite{Peres}). And in fact, the information $\Delta
I_{\Ens_M}(\tilde{M})$ from the reversion $\tilde{M}$ (now with {\em a
priori} probability given by the posterior probability $\ap_M$ from
the previous measurement $M$) is negative, and cancels exactly the
previous information $\Delta I_{\Ens}(M)$. However,
it is not always possible to erase the information from a measurement 
with another one, and, in common situations the information is
permanent, \ie it cannot be erased as in the case of a customary von
Neumann measurement. From the above considerations we learn the
general lesson: 1) in some cases the ``disturbance'' of two measurement
outcomes in cascade can be lower than that from a single measurement
outcome, since, at least, there are cases in which we can revert the
measurement---\ie with no overall disturbance---whence, more
generally, we can partially undo the disturbance from a previous
measurement; 2) when some disturbance is undone, then necessarily some
information is lost.  
\subsection{The case of measuring an observable}\label{s:obs}
When the quantum measurement is the measurement of an observable?
This is the case in which the positive operator valued measure (POVM)
of the measurement is commutative, namely the POVM is jointly
diagonalized on the same orthonormal basis, 
say $|x\>$. In fact, let's denote by $\{P_y\}$ with $P_y\ge 0$ and
$\sum_y P_y=I$ the POVM of the measurement. We can conveniently write
the joint diagonalization as follows
\begin{equation}
P_y |x\> =p(y|x)|x\>,
\end{equation}
where the eigenvalue $p(y|x)$ of $P_y$ on the eigenvector $|x\>$ is
denoted as a conditional probability, since we must have $p(y|x)\ge
0$, and $\sum_y p(y|x)=1$---and, in fact, we can interpret the
eigenvalue $p(y|x)$ as the conditional probability of getting $y$ when
the ``true'' value was $x$ instead. It is clear that the measurement
of an observable corresponds to our state-discriminating framework
when the input ensemble is the set of orthogonal states  $\{|
x\>\}$. A pure measurement that corresponds to the observable 
$X\doteq\{| x\>\}$ must be made of contractions $M_y$ with $M_y^\dag
M_y\equiv P_y$ with singular value decomposition $M_y=X_{M_y}\Sigma(M_y)
\Pi_y^\dag Y^\dag$ with right unitary operators $Y_y=Y \Pi_y $ giving
$Y_y^\dag |x\>=\Pi_y^\dag |n\>$, namely giving the same orthonormal
basis $\{|n\>\}$ on which $\Sigma(M_y)=\diag[\sigma_1(M_y),\sigma_2(M_y),
\ldots,\sigma_d(M_y)]$ is diagonal, apart from a permutation 
$\Pi_y$ of the basis $\{|n\>\}$. This is equivalent to say that the most general
form of the contraction $M_y$ is $M_y=W_y\sum_x\sqrt{p(y|x)}|x\>\<
x|$, with $W_y$ unitary: in other words, there is a unitary $W_y$
such that $[W_y^\dag M_y,|x\>\< x|]=0$ $\forall x$. 
The measurement is {\em complete}---\ie it scans the whole spectrum
$\sigma(X)\doteq \{x\}$ of the observable $X$ with
$|\sigma(X)|=d$---when $\rnk(M)=d$. The measurement is {\em non
degenerate}---namely each outcome $y$ corresponds unambiguously to a
unique most probable 
value $x$---if $\sigma_1(M_y)> \sigma_2(M_y)$, which means that
$p(y|x)$ for each $y$ has a non degenerate maximum versus $x$. The
optimal probability $p_{rev}(M_y)$ of reversing the contraction $M_y$
is given by $\sigma_d^2(M_y)$ and can be conveniently bounded as
$p_{rev}(M_y)\le [\prod_n\sigma_n^2(M_y)]^{1/d}$. Upon rewriting the
singular values in terms of the conditional probabilities and after
summing over all outcomes $y$ we get the bound for the average reversion
probability $\overline{p_{rev}}\le B(X:Y)$ where $B(X:Y)= \sum_y
[\prod_{x\in\sigma(X)}p(y|x)]^{1/|\sigma(X)|}$ is the Bhattacharyya
overlap bound derived in Ref. \cite{Ban}. We see that $0\le B(X:Y)\le
1$, with $B(X:Y)=0$ when $p(y|x)$ is vanishing for some values of
$x,y$, and $B(X:Y)=1$ when $p(y|x)$ is independent on $x$ for every
$y$. Therefore, the measurement has more chance of being reverted
---\ie it makes ``less disturbance''--- when the conditional
probability distribution is more ``flat'' versus $x$, namely the
information on $x$ is smaller.
\par The repeated application of a complete non degenerate measurement
of an observable $X$ provides another instructive example of the
information-disturbance trade-off. In fact, we can apply the
measurement many times on the same quantum system prepared in the
ensemble of orthogonal states $\{|x\>\}$, compensating the measurement
back-action with the conditional unitary transformation $W_y^\dag$. 
In this way we will make no disturbance
on the quantum system---which will always remain in its original
state---and, at the same time, from the statistics of the outcomes
we can also have perfect discrimination in the limit of infinitely
many repetitions. However, since a cascade made of more repetitions
will correspond to an overall conditioned probability more and more
sharply peaked around the ``right'' value $x$, the contraction
corresponding to the cascade will also have a decreasingly smaller chance
of reversion, and in the limit of infinite repetitions it will
approach a rank-one von Neumann measurement. Here we see that in
principle it is possible to extract perfect non erasable information even
by using a knowingly reversible measurement, however, performing the
measurement infinitely many times on the same quantum system. It is
clear that the information retrieved from the measurement on the input
state can be perfect only when the input ensemble is $\{|x\>\}$,
otherwise it will be lower than the maximum value (given by the Holevo
bound\cite{nielsenbook}), and, in particular, it is zero when the
input ensemble corresponds to the observable $Y$ ``conjugated'' to
$X$, namely the input states $\{|y_k\>,\,k=1,\ldots d\}$ are of the form
$|y_k\>=d^{-\frac{1}{2}}\sum_{l=0}^{d-1} e^{ikl2\pi/d}|x_l\>$ where
the spectrum of $X$ has been labeled with $x_l$.
\subsection{What is disturbance?}\label{whatis}
We cannot give a definition of disturbance that can be good for all
situations, since its definition must be suited to the particular
problem at hand. For example, a definition in terms of the
fidelity between input and output\cite{Fuchs-Peres} can be suited to
some quantum crypto-analysis: however, we cannot consider it as a
measure of the {\em in-principle} disturbance on the measured system,
since we would have disturbance also from a unitary transformation, which can
be reversed at will on any unknown input state. As another example,
when we want to account for the possibility of reversing the
measurement approximately by a unitary transformation, a suitable
definition of the disturbance $D(M)$ from a contraction $M$ should seize how
much the output $|\psi_M\>$ in Eq. (\ref{outpsi}) is {\em unitarily
uncorrelated} with the input $|\psi\>$, since we would say that
there is no disturbance if $|\psi\>$ and $|\psi_M\>$ are connected by
a fixed unitary transformation---say $V$---independently on 
$|\psi\>$. Then we would define the ``disturbance'' as $D(M)=1-C(M)$,
where $C(M)$ is the {\em input-output unitary correlation of $M$} defined as
the fidelity between $|\psi_M\>$ and $V|\psi\>$ for unitary $V$, averaged
over all $|\psi\>$ [with the joint probability $p(M,\psi)$], and
then maximized over $V$, namely
$C(M)=\max_{V}\overline{|\<\psi|V^\dag|\psi_M\>|^2}$.  A
straightforward calculation gives
$C(M)=\frac{1}{d(d+1)}[\N{M}_1^2+\N{M}_2^2]$. 
We can see that $C(M)$ approaches its maximum $C(M)=1$ for
contraction $M$ close to a unitary (all singular values approach 1), whereas it
is minimum $C(M)=2/d(d+1)$ for a rank-one $M$. Notice that here
$D(M)=1-C(M)$ is a Schur-convex function of the vector
$[\sigma_i^2(M)]$ of squared singular values of $M$. 
\par The ``disturbance''  $D(M)=1-C(M)$ sizes our inability of
approximately revert $M$ by a unitary transformation.  More generally, if
we want to define $D(M)$ in a way which is related to our ability
in-principle of reversing $M$, we must consider that reversion is
generally achieved by another measurement. Then, the definition of 
disturbance must satisfy the following requirements: 
\begin{enumerate}
\item \label{i1} The disturbance $D(M)$ due to $M$ must be a
function only of the probabilities of reversing its effect, not on how
the reversion is performed. Therefore, we must have $D(M)=D(UM)$,
for all unitary $U$, namely the disturbance is a function only of the
POVM element $M^\dag M$ of the measurement.
\item\label{i2} If we look for a definition of $D(M)$ which is a
property of $M$ only, independently on the input state, then 
in addition to the requirement \ref{i1} we must also have $D(M)=D(MV)$
for all unitary $V$. This means that the disturbance must be a
function of the singular values of $M$ only, namely
$D(M)=f(\{\sigma_l(M)\}).$ Therefore, our definition of $D(M)$ should
be of this form at least for the input universal ensemble.
\item\label{i3} We expect that the disturbance will be minimum for
unitary $M$, and maximum for $\rnk(M)=1$ (Gordon-Louisell 
measurement \cite{GordonLouisell}, \eg von Neumann): since in general
the definition of $D(M)$ should also depend on the input ensemble, these
two extreme cases at least should hold for the case of the input
universal ensemble. 
\end{enumerate}
\subsection{Majorization trade-offs.}
In the search for general trade-offs between ``information'' and
``disturbance'' for a quantum measurement at the single-outcome level
we will try to accomplish the following aim. While satisfying the
above requirements \ref{i1}-\ref{i3}, we look for general
inequalities which will guarantee the trade-off independently 
on the specific quantities that will be used for both ``information''
and ``disturbance'', to be suited to the particular problem at hand.
Notice that the usual information in Eq.  (\ref{infosingle}) is the
sum of two contributions, of which the first one $H(\Ens)$ is
independent on $M$, whereas the second $-H(\Ens_M)$ 
is a Schur convex function of the conditioned
probabilities $a(M|\psi)$. Therefore, if we want our trade-off to be
true also for the usual information (\ref{infosingle}), we should look
for a majorization relation $\ap_M\prec\z_M$ between the vector
$\ap_M=[a(\psi|M)]$ and a vector $\z_M=\z(\sigma_i(M),\Ens)$ having
components that depend on the singular values $\sigma_i(M)$ of $M$
along with quantities related to the ensemble $\Ens$, and such that for
the input universal ensemble will be a function of
$\sigma_i(M)$ only [for majorization theory see
Ref. \cite{Nielsen,Marshall}]. This will guarantee the 
trade-off by just taking for $\z_M$ the same Schur-convex function
$f=-H(\ap_M)$ that we have in the information, namely  
$f(\z_M)\equiv-H(\z_M)$. Moreover, the majorization relation will guarantee the
trade-off for any other choice of Schur-convex function, depending on
the problem, in which the ``information'' is a function of $\ap_M$,
and the ``disturbance'' is the same function of $\z_M$. Notice, however, that the power
of the majorization approach, is also its weakness. In fact, since a
majorization relation will guarantee the trade-off for all Schur-convex
functions, it may be possible that for a given function ($f=-H$ in
our case) the trade-off could be true more generally than 
for $\ap_M\prec\z_M$. Finally, we want to emphasize that the convexity of
the function $f$ is unrelated with the assertion that ``the
disturbance from a set of $M$ randomly chosen is always lower than
their averaged disturbance'', since in our case the definition of
disturbance is given only for pure contractions, as we are
concerned only with pure measurements. On the other hand, as we will
see in the following, when we consider the complete measurement with
all possible outcomes, we can easily average the trade-off over the
outcomes with their probabilities of occurrence.
\par Looking for a majorization relation involving $\ap_M$ is
equivalent to look for a majorization relation for the joint
probabilities $a(M,\psi)$, since the two are related by a fixed
normalization constant given by 
the overall probability $p_\Ens(M)$ of occurrence of $M$. It is
easy to derive a weak majorization relation as follows
\begin{equation}
a(M,\psi_j)=a(\psi_j)a(M|\psi_j)=a(\psi_j)
\<\psi_j|Y_M\Sigma_M^2Y_M^\dag|\psi_j\>=\sum_{i=1}^d \sigma_i^2(M)
a(\psi_j)|\< i|Y_M^\dag|\psi_j\>|^2\doteq
\sum_{i=1}^d W_{j i} \sigma_i^2(M),\label{infostochast}
\end{equation}
where $M=X_M \Sigma_M Y_M^\dag$ is the singular value decomposition of
$M$, and $\{|i\>\}$ is an orthonormal basis on which $\Sigma_M$
has the canonical diagonal form. The rectangular matrix $W_{j i}\doteq
a(\psi_j)\< i|Y_M^\dag|\psi_j\>|^2$ is double sub-stochastic,  
since $\sum_i W_{j i}=a(\psi_j)\Tr[Y_M^\dag|\psi_j\>\<\psi_j| Y_M]=
a(\psi_j)$, and $\sum_j W_{j i}=\< i|Y_M^\dag\rho_\Ens Y_M|i\>\le 1$. This means
that the following weak majorization relation (symbol $\prec_w$) holds
\begin{equation}
[a(M,\psi_j)]\prec_w [\sigma_i^2(M)].\label{majw}
\end{equation}
However, the weak majorization relation $\prec_w$ will guarantee
trade-offs for a choice of Schur-convex function that is also
increasing on its domain\cite{Marshall} [again, this does not mean
that the trade-off cannot hold for some particular Schur-convex function].
\par A majorization relation between the vector $[a(M,\psi_j)]$ and
a vector containing the singular values of $M$ can be obtained by
expanding the probability $a(M,\psi_j)$ as follows
\begin{equation}
a(M,\psi_j)=a(\psi_j)\<\psi_j|Y_M\Sigma_M^2 Y_M^\dag|\psi_j\>=
\sum_i a(\psi_j)|\<\psi_j|Y_M|i\>|^2\sigma_i^2(M)=\sum_i S_{j
i}\lambda_i\sigma_i^2(M),\label{M1}
\end{equation}
where
\begin{equation}
\lambda_i=\< i|Y_M^\dag\rho_\Ens Y_M|i\>,\quad
S_{j i}=a(\psi_j)|\<\psi_j|Y_M|i\>|^2\lambda_i^{-1}.\label{S}
\end{equation}
Notice that $\lambda_i=\sum_j a(\psi_j)|\< i|Y_M^\dag|\psi_j\>|^2$ and
$\lambda_i=0$ if and only if $|\<\psi_j|Y_M|i\>|^2= 0,\forall j$, and
the sum in Eq. (\ref{M1}) is extended only to those terms for which
$\lambda_i>0$---say for $i=1,\ldots r\le\rnk(M)$. It follows that 
the $|\Ens|\times r$ matrix $S$ has the following rows and column sums
\begin{equation}
\sum_i S_{j i}=a(\psi_j)\<\psi_j|Y_M \zeta^{-1} 
Y_M^\dag|\psi_j\>\doteq s_j,\qquad\sum_j S_{j i}=1, 
\end{equation}
where $\zeta=\sum_i \lambda_i |i\>\< i|$. Notice that generally
$\zeta\neq\rho_\Ens$ and we have $\zeta=\rho_\Ens$ when 
$\rho_\Ens$ is diagonal with $M^\dag M$, namely when $[\rho_\Ens,
M^\dag M]=0$, in which case we are guaranteed that
$s_j\le 1,\forall j$, whereas in general $s_j$ can be greater than
unit. We will call the ensemble $\Ens$ {\em parallel} to $M$ when
$\rho_\Ens$ commutes with $M^\dag M$, and  {\em quasi-parallel} to $M$
when $s_j\le 1,\forall j$. Ensembles that are parallel to any $M$ are
obviously the maximally chaotic ones, for which $\rho_\Ens=d^{-1}I$.
For ensembles quasi-parallel to $M$ the $|\Ens|\times r$ matrix $S$ in
Eq. (\ref{S}) can be augmented to a $(|\Ens|+r)\times (|\Ens|+r)$
stochastic matrix as follows    
\begin{equation}
\tilde{S}=\matrix{
\fbox{\parbox[t][2cm][c]{1cm}{\begin{center}$S$\end{center}}}
\!\!\! & \!\!\! 
\fbox{\parbox[t][2cm][c]{2cm}{\begin{center}$\diag\{1-s_j\}$\end{center}}} & \cr
\fbox{\parbox[t][1cm][c]{1cm}{\begin{center}$0$\end{center}}}
\!\!\! & \!\!\! 
\fbox{\parbox[t][1cm][c]{2cm}{\begin{center}$\transp{S}$\end{center}}}
& \cr}\label{Smatrix}
\end{equation}
By padding the vectors $[a(M,\psi_j)]$ and $[\lambda_i\sigma_i^2(M)]$
with $r$ and $|\Ens|$ additional zeros, respectively, 
Eqs. (\ref{M1}) and (\ref{Smatrix}) guarantee the following
majorization relation 
\begin{equation}
[a(M,\psi_j)]\prec [\lambda_i\sigma_i^2(M)],
\end{equation}
and upon normalizing both vectors we have
\begin{equation}
\ap_M\prec {\mathbf z_M},\label{majz}
\end{equation}
with
\begin{equation}
(\z_M)_i=p^{-1}_\Ens(M)\lambda_i\sigma_i^2(M).\label{zM}
\end{equation}
For ensembles $\Ens$ that are not quasi-parallel to $M$
we can always build a {\em squashed} ensemble $\tilde{\Ens}$ that is
quasi-parallel to $M$ by replicating the state $|\psi_j\>$
corresponding to $s_j>1$ in sufficiently many identical copies
$|\psi^{(j)}_l\>\equiv |\psi_j\>$ distributed with 
probabilities $a(\psi^{(j)}_l)=q^{(j)}_l a(\psi_j)$, with $\sum_l 
q^{(j)}_l=1$, such that $s_j\max\{q^{(j)}_l\}\le 1$.
\subsection{Information disturbance trade-offs}
From Eq. (\ref{majz}) it follows that for ensembles quasi-parallel to
$M$ we have $-H(\ap_M)\le -H(\z_M)$, and for the information
on which-state retrieved from the occurrence of $M$ we have
\begin{equation}
\Delta I_\Ens(M)\le H(\Ens)-H(\z_M).\label{Iz}
\end{equation}
If the ensemble is not quasi-parallel to $M$, by considering any
squashed ensemble $\tilde{\Ens}$  we obtain
\begin{equation}
\Delta
I_\Ens(M)\le H(\Ens)-H({\mathbf
z}_M)-\sum_j[a(\psi_j)-p(\psi_j|M)]H({\mathbf q}^{(j)}),\label{stuff}
\end{equation}
but, unfortunately, the last quantity in Eq. (\ref{stuff}) has no
definite sign. For this reason, in the following we will focus
attention only on ensembles that are quasi-parallel to $M$. 
\par When considering a complete pure measurement ${\cal
M}=[M_1,M_2,\ldots 
M_n]$ with $\sum_i M^\dag_i M_i= I$ we can average both sides of
Eq. (\ref{Iz}) on outcomes $i$ with the probability of occurrence
$p_\Ens(M_i)$, and obtain 
\begin{equation}
\Delta I_\Ens({\cal M})\le H(\Ens)-\< H(\z_{M_i})\> ,\label{Izav}
\end{equation}
where $\<\ldots\>$ denotes the averaging over outcomes $i$. The
quantity $-H(\z_M)$ can be regarded as a kind of
``disturbance'' due to $M$. Notice that 
\begin{equation}
-\log\rnk(M)\le -H(\z_M)\le 0,
\end{equation}
The disturbance is minimum when $\sigma_i^2(M)\propto\lambda_i^{-1}$,
and maximum for rank-one $M$ (Gordon-Louisell measurements) or when
there is only one right-vector $Y|i\>$ of $M$ in the range of
$\rho_\Ens$. Notice that for general ensemble the disturbance is not
minimum for unitary $M$: this is a phenomenon due to the
occurrence of negative informations analyzed previously, \eg a
measurement reverting a previous one undoes its disturbance, namely it
makes ``less disturbance'' than a unitary transformation. In
particular, when the ensemble is orthogonally split by $M$ and
$\Ens_M^\parallel$ is itself orthogonal, then the minimum
disturbance will be exactly equal to the information gain
$-H(\Ens_M^\parallel)$ in Eq. (\ref{revinf}) from a successfully
reverted measurement. For orthogonal ensembles (generally not split) 
we have  in average over outcomes
\begin{equation}
\Delta I_\Ens({\cal M})\le S(\rho_\Ens)-\< H(\z_M)\>\le
\chi(\Ens),\label{hol}
\end{equation}
where $S(\rho)=-\Tr[\rho\log\rho]$ denotes the von Neumann
entropy, and $\chi(\Ens)=S(\rho_\Ens)-\sum_j a_j S(\rho_j)$ is the
Holevo bound for the ensemble with density operator 
$\rho_\Ens=\sum_j a_j\rho_j$ for {\em a priori} probabilities and
states $a_j$ and $\rho_j$, respectively. Eq. (\ref{hol}) gives a bound
for the information retrieved from the single outcome that is tighter
than Holevo bound [in our case the {\em a priori} states
$\rho_j=|\psi_j\>\<\psi_j|$ are pure, and
$\chi(\Ens)=S(\rho_\Ens)$]. The information disturbance trade-off
(\ref{hol}) asserts that we can make less
disturbance at the price of retrieving less information than the
available one. Also notice that in the present case of orthogonal
input ensemble a measurement ${\cal M}$ made of random unitary
transformations will give minimum disturbance and zero information.
\par
\begin{figure}[hbt]
\begin{center}
\epsfxsize=.3\hsize\leavevmode\epsffile{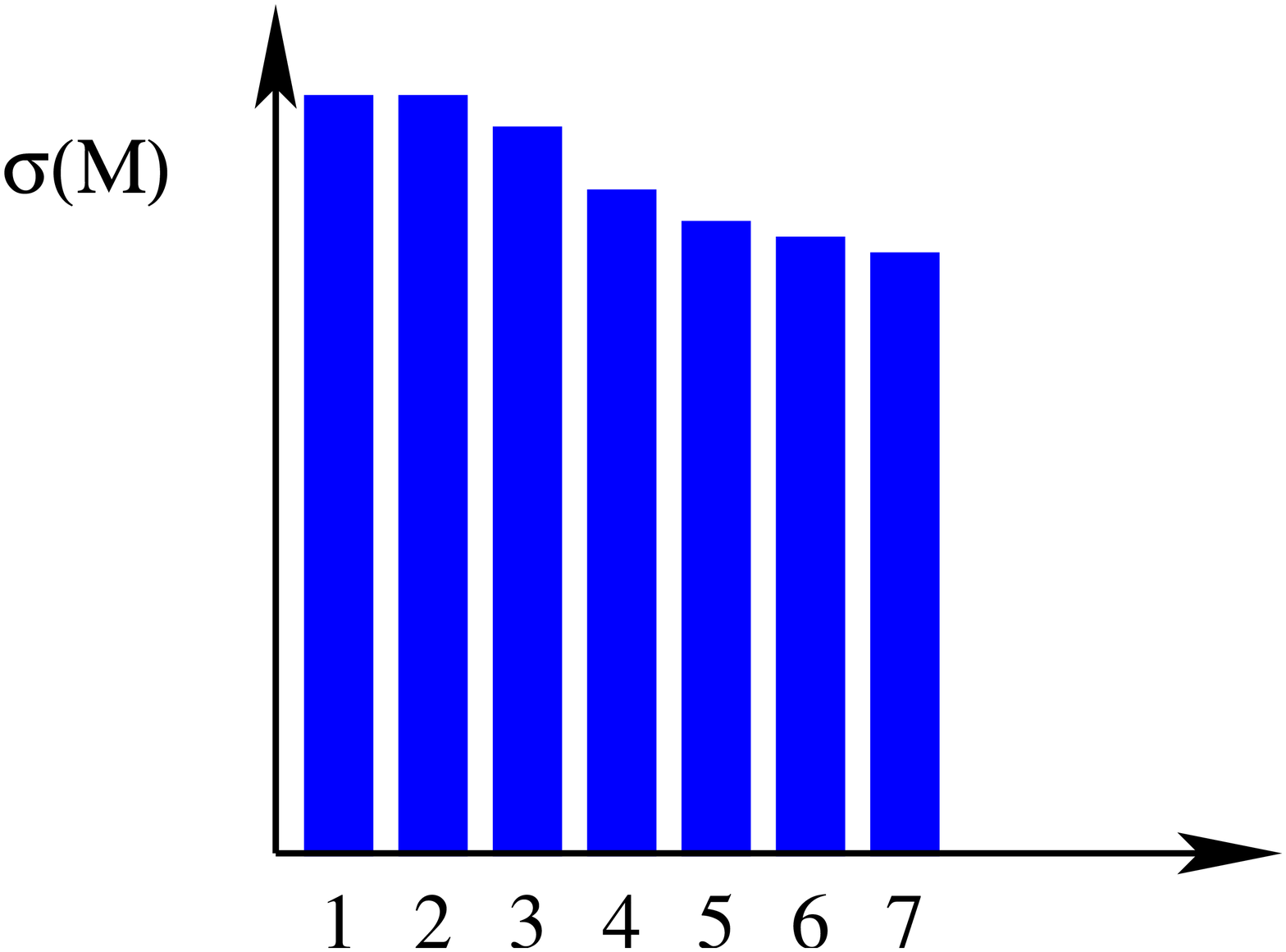} 
\epsfxsize=.3\hsize\leavevmode\epsffile{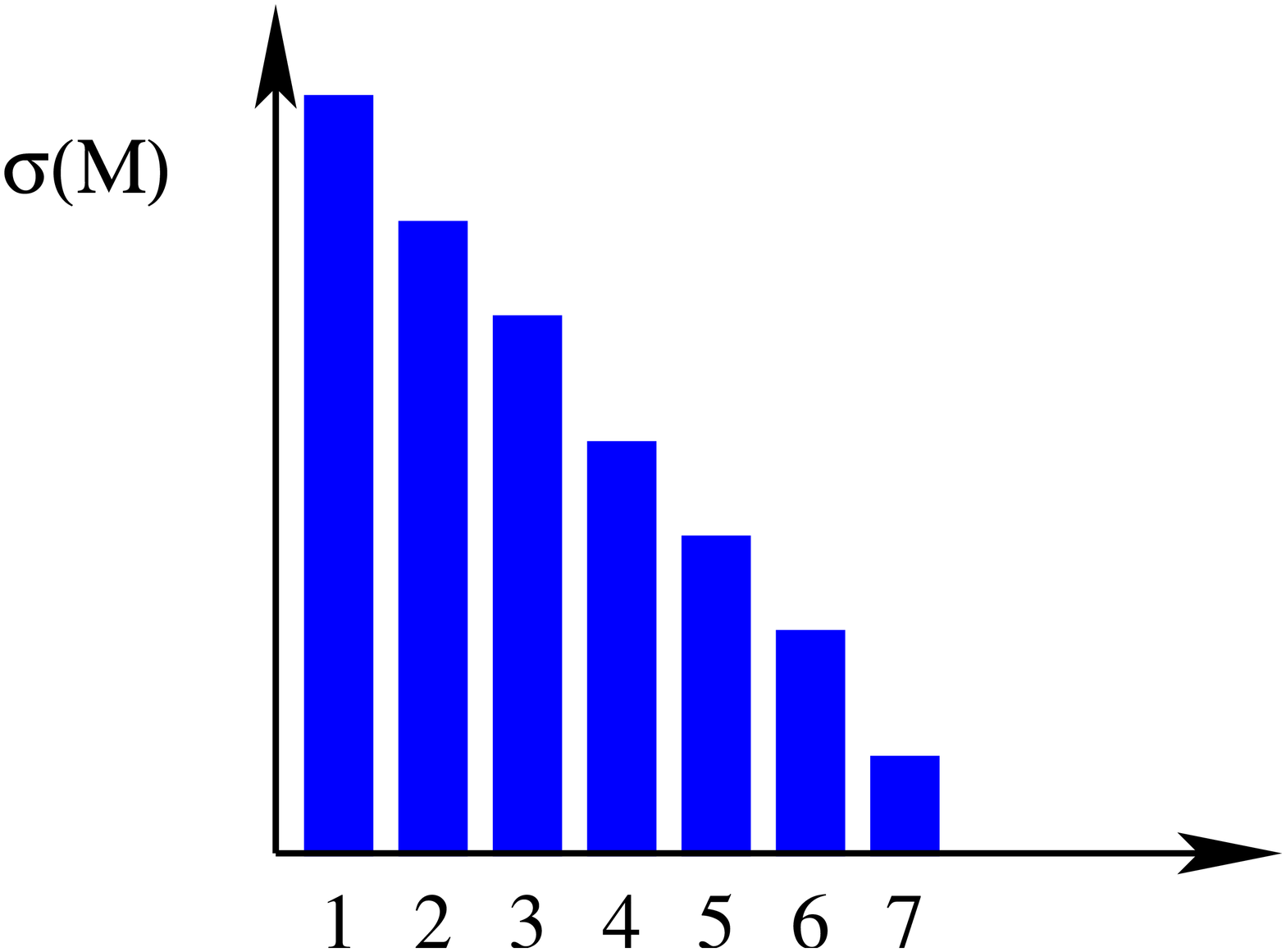} 
\end{center}
\caption{More and less disturbing measurement contraction $M$ (for 
input universal ensemble): the less disturbing $M$ (on the left) has
``more flat'' singular values.}\label{f:M}
\end{figure}

 We want to focus now on the simplest case in which the
ensemble $\Ens$ is parallel to $M$. Here we have
\begin{equation}
-H(\z_M)=-S((\rho_\Ens)_M),\label{vnM}
\end{equation}
namely our disturbance is equal to the opposite of the von Neuman
entropy of the ``reduced'' density operator $(\rho_\Ens)_M$ 
\begin{equation}
(\rho_\Ens)_M=\frac{M\rho_\Ens M^\dag}{\Tr[M\rho_\Ens M^\dag]}.\label{parallel}
\end{equation}
From Eqs. (\ref{vnM}) and (\ref{parallel}) we also see that for
ensembles parallel to $M$ our ``disturbance'' is also exactly equal to
the ``reduction of entanglement'' that $M$ would produce locally on any
entangled state $|\Psi\>$ that is a purification of $\rho_\Ens$,
namely, for
\begin{equation}
|\Psi_M\>\doteq \frac{M\otimes I |\Psi\>}{\N{M\otimes I |\Psi\>}},\qquad
\Tr_2[|\Psi\>\<\Psi|]=\rho_\Ens
\end{equation}
we will have $-H(\z_M)=-S(\Tr_2[|\Psi_M\>\<\Psi_M|])$. Notice that in
general,  $M$ can also probabilistically increase the entanglement of
$|\Psi\>$: this situation corresponds to the occurrence of negative
informations mentioned above, with disturbance less than that from a
unitary transformation. In the special case in which the ensemble is
also maximally chaotic (\eg for the universal ensemble), our
disturbance will be given by   
\begin{equation}
-H(\z_M)=\sum_i \frac{\sigma_i^2}{\N{M}_2} \log
\frac{\sigma_i^2}{\N{M}_2},\label{lastD}
\end{equation}
and the less disturbing is $M$, the ``more flat'' are its singular
values, with the largest mutual information being achievable with
rank-one measurements. This situation is depicted in
Fig. \ref{f:M}. From Eq. (\ref{lastD}) we see that our disturbance
``interpolates'' the definition of disturbance $D(M)=-\log\rnk(M)$
proposed by Ozawa \cite{ozawaunpub} for the trade-off $I(X|\rho)\le
S(\rho) -\log\rnk(M_x)$ for the ``information gain'' $I(X|\rho)\doteq
S(\rho)-\sum_x p(x|\rho)S(\rho_x)$ \cite{ozawajmp} from a pure quantum
measurement made of contractions $M_x$ all with the same rank
$\rnk(M_x)$, with $\rho$ the input state,
$p(x|\rho)=\Tr[M_x^\dag M_x\rho]$, and $\rho_x=M_x\rho
M_x^\dag/\Tr[M_x^\dag M_x\rho]$ the output state. 
\par We conclude this section by noticing that the present definition
of disturbance explains the information-disturbance trade-off in
quantum teleportation\cite{bdms}, between the Alice's information on the
transmitted state and the disturbance at Bob on the received state, the
trade-off being tuned by switching on-off the entanglement of the
shared resource. Indeed, it is easy to see that in any teleportation
scheme in which Alice performs a generic Bell measurement \cite{bdms}, 
the disturbance is just the opposite of the entanglement of the state
$|\Psi\>$ of the shared resource.  
\section{Concluding remarks}
In this paper we have considered an ideal in-principle quantum
measurement at the single-outcome level, which is then described by a
single contraction. We have analyzed the possibility of measurements
that are knowingly reversible, showing that measurement reversion
necessarily erases the information from the reverted measurement. This
also clarifies that it is possible in principle to undo the effect of a measurement,
however, at the expense of losing some previously retrieved
information. Information-disturbance trade-offs have been presented,
where the ``disturbance'' depends on the probabilities of reverting 
the measurement.  Two majorization relations have been given:
the weak majorization (\ref{majw}), which holds for any ensemble, and
the majorization (\ref{majz}), which hold for ensembles ``quasi-parallel''
to the measurement contraction $M$. These relations represent
trade-offs that are independent on the particular analytical form of
information and disturbance. When considering the customary
mutual information, the majorization (\ref{majz}) leads us to consider
the quantity $-H(\z_M)$ as a ``disturbance'', with the vector $\z_M$
depending on the singular values of $M$ and on the input ensemble 
$\Ens$ as given in Eq. (\ref{zM}).  Such quantity satisfies all the
requirements that we gave for a general disturbance, and behaves as 
expected in all known cases. Even though the information-disturbance
trade-off (\ref{Iz}) has been proved for ensemble quasi-parallel to $M$
(since it has been derived from the majorization relation (\ref{majz})) 
Eq. (\ref{Iz}) can have a more general validity, and an alternative
derivation will be the subject of a forthcoming work.
\section*{\protect{\noindent\normalsize\bf Acknowledgments}}
I acknowledge illuminating discussions with M. Ozawa. I'm also
greateful to E. Giannetto for providing me some relevant historical
references, and for interesting conversations. Finally I'm greateful
to M. Sacchi and P. Lo Presti for careful reading the manuscript.
This work has been founded by the EC program ATESIT, Contract
No. IST-2000-29681, and by DARPA Grant No. F30602-01-2- 0528.
\section*{\normalsize\bf References}
\begin{description}\itemsep=0pt
\bibitem[1]{nielsenbook}  I. L. Chuang and M. A. Nielsen, {\em Quantum Information and
Quantum Computation}, Cambridge University Press (Cambridge UK 2000). 
\bibitem[2]{Heisenberg} W. Heisenberg, Zeitschrift  f\"ur Phisik {\bf 43} 172-198 (1927).  
\bibitem[3]{Heisenberg_book} W. Heisenberg, {\em The physical
principles of Quantum Theory}, Univ. Chicago Press, Chicago (1930) - Dover NY.  
\bibitem[4]{Ruark} A. E. Ruark, Bull. APS, {\bf 2} 16 (1927); Phys. Rev. {\bf 31} 311-312 (1928).
\bibitem[5]{vonneuman} J. von Neumann, {\em Mathematical Foundation
of Quantum Mechanics}, Princeton University Press, Princeton N. J. (1955).   
\bibitem[6]{Messiah} A. Messiah, {\em Quantum Mechanics},
North-Holland Phys. Publ. (Amsterdam 1986).
\bibitem[7]{Braginski} V. B. Braginsky\u{\i}i and Yu. I. Vorontsov,
Sov, Phys.-Usp., {\bf 17} (1975). 
\bibitem[8]{Caves} C. M. Caves, K. S. Thorne, R. W. P. Drever,
V. D. Sandberg, and M. Zimmermann, Rev. Mod. Phys. {\bf 52}, 341 (1980).  
\bibitem[9]{Yuen_contractive} H. P. Yuen, Phys. Rev. Lett. {\bf 51}, 719 (1983). 
\bibitem[10]{Lynch} R. Lynch, Phys. Rev. Lett. {\bf 52}, 1730 (1984);
see also Yuen's response \cite{Yuen_SQL_unpub}.
\bibitem[11]{Caves-defense} C. M. Caves, Phys. Rev. Lett. {\bf 54}, 2465 (1985).
\bibitem[12]{Ozawa_SQL_break} M. Ozawa, Phys. Rev. Lett. {\bf 51}, 719 (1983). 
\bibitem[13]{Ozawa_SQL_break2} M. Ozawa, in {\em Squeezed and
Nonclassical Light}, ed. by P. Tombesi and E. R. Pike, Plenum, New
York 1989, pag. 263.
\bibitem[14]{Ozawa_model_SQL} M. Ozawa, Phys. Rev. A {\bf 41}, 1735 (1990). 
\bibitem[15]{Yuen_SQL_unpub} H. P. Yuen, {\em Violation of the
Standard Quantum Limit by Realizable Quantum Measurements}
(unpublished)
\bibitem[16]{Braginski_book} V. B. Braginsky and F. Ya. Kalili,
{\em Quantum measurement}, Ed. by. Kip. S. Thorne, 
Cambridge University Press, Cambridge G. B. (1992).
\bibitem[17]{Landau} L. D. Landau and E. M. Lifshitz, {\em Quantum
Mechanics}, Pergamon, Oxford (1965).
\bibitem[18]{Peres} A. Peres, {\em Quantum theory: concepts and
methods}, Kluwer, Dordrecht, (1993). 
\bibitem[19]{Jammer2} Max Jammer, {\em The conceptual development of
quantum mechanics}, Mc Graw-Hill, NY (1966). 
\bibitem[20]{Ozawa_reduct} M. Ozawa, Phys.\ Lett.\ A {\bf 282} 336 (2001).
\bibitem[21]{Muynck} W. M. de Muynck, Found. of Phys. {\bf 30} 205-225 (2000). 
\bibitem[22]{Bohm} D. Bohm, {\em Quantum Theory}, Dover, Mineola N. Y. (1989).
\bibitem[23]{ArthursKelly} E. Arthurs and J. L. Kelly, Bell. Syst. Tech. J.,  {\bf 44} 725-729 (1965). 
\bibitem[24]{GordonLouisell} J. P. Gordon and W. H. Louisell, in {\em Physics 
of Quantum Electronics}, pp. 833-840, McGraw-Hill, (New York, 1966).
\bibitem[25]{Yuen_joint} H. P. Yuen, Phys. Lett. {\bf 91A}, 101 (1982).
\bibitem[26]{Goodman} E. Arthurs and M. S. Goodman, Phys. Rev. Lett.
{\bf 60} 2447 (1988).
\bibitem[27]{Lamb} W. E. Lamb Jr., Physics Today {\bf 22}A 23 (1969).
\bibitem[28]{Jammer1}Max Jammer, {\em The Philosophy of Quantum
Mechanics}, Wiley, NY (1974).
\bibitem[29]{Beller} M. Beller, {\em Quantum Dialogue}, University of
Chicago Press, Chicago (1999).
\bibitem[30]{Roychoudhury} Chandrasekhar Roychoudhury, Found Phys. {\bf 8} 845 (1978).
\bibitem[31]{Uffink} J. Hilgevoord and J. B. M. Uffink, in {\em Sixty-two Years of Uncertainty},
ed. by A. I Miller, Plenum, NY (1990).
\bibitem[32]{Robertson} H. P. Robertson, Phys.Rev. {\bf 34}, 163-164
(1929); Phys. Rev. {\bf 35}, 667-667 (1930); Phys. Rev. {\bf 46}, 794-801 (1934).  
\bibitem[33]{Condon} E. U. Condon, Science LXIX 573 (1929).
\bibitem[34]{Schroedinger} E. Schr\H{o}dinger,  Sitz. Preus. Acad. Wiss. (Phys.-Math. Klasse),
{\em 19}, 296-303 (1930). 
\bibitem[35]{Deutsch} D. Deutsch, Phys. Rev. Lett. {\bf 50} 631 (1983).
\bibitem[36]{Massen} H. Massen and J. B. M. Uffink, Phys. Rev. Lett. {\bf 60} 1103 (1988).
\bibitem[37]{Hall} M. J. W. Hall, Phys. Rev. A {\bf 49} 42 (1994). 
\bibitem[38]{Sudarshan} E. C. G. Sudarshan, {\em Paraxial optics and higher
uncertainties} ICSSUR Napoli (1999 ).
\bibitem[39]{DArianoYuen_clon} G. M. D'Ariano and H. P. Yuen,
Phys. Rev. Lett. {\bf 76} 2832 (1996). 
\bibitem[40]{Yamamoto} O. Alter, and Y. Yamamoto, Phys. Rev. Lett. {\bf 74}, 4106 (1995).
\bibitem[41]{Aharonov} Y. Aharonov, J. Anandan, L. Vaidman,
Phys. Rev. A {\bf 47}, 4616 (1993); Y. Aharonov and L. Vaidman, Phys. Lett. A {\bf 178}, 38 (1993).
\bibitem[42]{Kitagawa} M. Ueda and M. Kitagawa, Phys. Rev. Lett. {\bf 68}, 3424 (1992).
\bibitem[43]{Imamoglu} A. Imamoglu, Phys. Rev. A {\bf 47}, R4577 (1993).
\bibitem[44]{Royer} A. Royer, Phys. Rev. Lett. {\bf 73} 913 (1994). 
\bibitem[45]{Wootters_clon} W. K. Wootters, W. H. Zurek, Nature {\bf
299},  802 (1982).
\bibitem[46]{Yuen_clon} H. P. Yuen, Phys.\ Lett.\ A{\bf 113} 405 (1986).  
\bibitem[47]{Fuchs-Peres} C.~A. Fuchs and A.~Peres, Phys. Rev. A, {\bf 53} 2038 (1996).
\bibitem[48]{Fuchs} C. A. Fuchs, Fortschr. Phys. 46 535-565 (1998).
\bibitem[49]{Barnum} H. Barnum, report University of Bristol, (2000). 
\bibitem[50]{Banaszek} K. Banaszek, Phys. Rev. A{\bf 64} 052307 (2001)
\bibitem[51]{ozawaunpub} M. Ozawa, (private communication).
\bibitem[52]{Belavkin} V. P. Belavkin, Progr. Quant. Electr. {\bf 25}
1 (2001)
\bibitem[53]{Nielsen} M. A. Nielsen, Phys. Rev. Lett. {\bf 83} 436-439
(1999). 
\bibitem[54]{Marshall} A. W. Marshall and I. Olkin, {\em
Inequalities: Theory of Majorization and its Applications}, Academic
Press, N. Y. (1979).
\bibitem[55]{Ban} J. Phys. A: Math. Gen. {\bf 34} 1 (2001).
\bibitem[56]{Royer_errata} A. Royer, Phys. Rev. Lett. {\bf 74} 1040 (1995)
[Errata suggested by J. Finkelstein, B. Huttner, and N. Gisen].
\bibitem[57]{ozawajmp} M. Ozawa, J. Math. Phys. {\bf 27} 759 (1986).
\bibitem[58]{bdms} S. L. Braunstein, G. M.\ D'Ariano, G.\ J.\
Milburn, and M. F. Sacchi, Phys. Rev. Lett. {\bf 84} 3486 (2000).  
\end{description}
\end{document}